\documentclass[12pt]{article}
\usepackage{epsf}
\renewcommand{\theequation}{\arabic{equation}}
\newcommand{\EQ}{\begin{equation}}

\newcommand{\EN}{\end{equation}}

\newcommand{\bear}{\begin{eqnarray}}
\newcommand{\ear}{\end{eqnarray}}

\begin{document}

\topmargin 0pt
\oddsidemargin 5mm
\newcommand{\NP}[1]{Nucl.\ Phys.\ {\bf #1}}
\newcommand{\PL}[1]{Phys.\ Lett.\ {\bf #1}}
\newcommand{\NC}[1]{Nuovo Cimento {\bf #1}}
\newcommand{\CMP}[1]{Comm.\ Math.\ Phys.\ {\bf #1}}
\newcommand{\PR}[1]{Phys.\ Rev.\ {\bf #1}}
\newcommand{\PRL}[1]{Phys.\ Rev.\ Lett.\ {\bf #1}}
\newcommand{\MPL}[1]{Mod.\ Phys.\ Lett.\ {\bf #1}}
\newcommand{\JETP}[1]{Sov.\ Phys.\ JETP {\bf #1}}
\newcommand{\TMP}[1]{Teor.\ Mat.\ Fiz.\ {\bf #1}}
     
\renewcommand{\thefootnote}{\fnsymbol{footnote}}

\setcounter{page}{0}
\begin{titlepage}     
\begin{flushright}
UFSCAR-TH-02 
\end{flushright}
\vspace{0.5cm}
\begin{center}
\large{ Yang-Baxter equation for the asymmetric eight-vertex model.
 } \\
\vspace{1cm}
\vspace{1cm}
{\large W. Galleas and M.J. Martins} \\
\vspace{1cm}
\centerline{\em Departamento de F{\'{\i}}sica, Universidade Federal de S\~ao Carlos}
\centerline{\em Caixa Postal 676, 13565-905, S\~ao Carlos, Brazil}
\vspace{1.2cm}   
\end{center} 
\begin{abstract}
In this note we study \`a la Baxter \cite{BA} the possible integrable
manifolds of the asymmetric eight-vertex model. As expected they occur
when the Boltzmann weights are either symmetric or satisfy the
free-fermion condition but our analysis clarify the reason both manifolds
need to share a universal invariant. We also show that the free-fermion
condition implies three distinct classes of integrable models.
\end{abstract}
\vspace{.2cm}
\vspace{.2cm}
\centerline{July 2002}
\end{titlepage}

\renewcommand{\thefootnote}{\arabic{footnote}}

Exactly solved vertex models play a fundamental role in classical
statistical mechanics. The most important of these is the 
so-called eight-vertex model which contains as special cases
most systems on a plane square lattice \cite{BA}. The general
asymmetric eight-vertex model possesses six different Boltzmann
weights $a_{\pm},b_{\pm},c $ and $d$ whose transfer matrix
can be written as
\begin{equation}
T= \Gamma r_2[ {\cal{L}}_L \cdots {\cal{L}}_1 ]
\end{equation}
where the trace is over the ordered product
of local operators ${\cal{L}}_j$ which are given by the following $2 \times 2$
matrix
\begin{equation}
{\cal{L}}_j=\left(\matrix{ 
a_{+}\sigma_j^{+}\sigma_j^{-}+b_{+} \sigma_j^{-}\sigma_j^{+} & 
d\sigma_j^{+}+c\sigma_j^{-} \cr
c\sigma_j^{+}+d\sigma_j^{-} & 
b_{-}\sigma_j^{+}\sigma_j^{-}+a_{-} \sigma_j^{-}\sigma_j^{+}  \cr } \right)
\end{equation}
and $\sigma_j^{\pm}$ are Pauli matrices 
acting on the sites $j$ of
an one-dimensional lattice.
The asymmetric eight-vertex model is known to be 
solvable in the manifolds
\begin{equation}
F(a_{\pm},b_{\pm},c,d)=0,
~ \frac{cd}{a_{+}b_{-}+a_{-}b_{+}}= I_1^{F},~
\frac{a_{+}^2+b_{-}^2-a_{-}^2-b_{+}^2}{a_{+}b_{-}+a_{-}b_{+}}=I_2^{F}
\end{equation}
\begin{equation}
a_{\pm}=b_{\mp},~~c=d~~~or~~~a_{\pm}=b_{\pm},~~c=d
\end{equation}
\begin{equation}
a_{+}=a_{-},~ b_{+}=b_{-},~ \frac{cd}{a_{+}b_{+}}= I_1^{B},~
\frac{F(a_{\pm},b_{\pm},c,d)}{a_{+}b_{+}}=I_2^{B}
\end{equation}
where $F(a_{\pm},b_{\pm},c,d)=a_{+}a_{-}+b_{+}b_{-}-c^2-d^2$ and
$I_{1,2}^{F,B}$ are arbitrary constants.

The manifold (3) is the so-called free-fermion model whose free-energy
was first calculated by Fan and Wu \cite{FW} and later re-derived
by Felderhof \cite{FE} who devised a method to diagonalize the 
corresponding transfer matrix. The integrability of the free-fermion
manifold is usually assumed from the fact that its transfer matrix
commutes with the $XY$ Hamiltonian as shown by Krinsky \cite{KI}
who used  a procedure first developed by Sutherland \cite{SU}.
Later on Barouch \cite{BO} and Kasteleyn \cite{KA} have revisited the problem of commuting asymmetric
eight-vertex transfer matrices and generalized Heisenberg Hamiltonians which leaded
Kasteleyn \cite{KA} to point out the existence of the manifolds (4). As stressed by this author,
however, such manifolds are trivial because they can be seen as set of independent 
one-dimensional models and therefore they should be disregarded.
On the other hand, the solution of the symmetric manifold was found
by Baxter through quite general approach
denominated commuting transfer matrix method which culminated
in the famous ``star-triangle'' relations \cite{BA}.
The fact that the symmetric eight-vertex transfer matrix commutes with
a related $XYZ$ Hamiltonian \cite{SU} has then been made more
precise because the latter is essentially a logarithmic derivative
of the former \cite{BA}. 

It would be quite desirable to extend the Baxter's method to the
asymmetric eight-vertex model and to rederive the manifolds
(3) and (5) from a unified point of view. 
Since this approach does not assume a priori the existence of a specific local
form for the corresponding Hamiltonian it can lead us 
to new integrable manifolds not covered
by the analysis of Barouch \cite{BO} and Kasteleyn \cite{KA}.
We recall that
much of the work on this problem, see e.g. refs. \cite{WA,SO,CH},
has been concentrated to analyze the Yang-Baxter equations
directly in terms of spectral parameters. Though this is
a valid approach it often hides the general integrable
manifolds in terms of specific parameterizations which need
to be found by a posteriori guess-work.
A more direct way would be first to determine the solvable
manifolds by an algebraic study \`a la Baxter 
of the corresponding
``star-triangle'' equations and afterwards to parameterize
them by using the theory of uniformization of biquadratic
polynomials \cite{BA}. 
It appears that Kasteleyn \cite{KA} was the first to make an effort toward
such analysis but the best he could do was to guess the manifold (3)
from known results by Felderhof besides clarifying the origin of the 
pseudo one-dimensional manifold (4)
as the linearization of the Yang-Baxter equation around a non-identity 
$4 \times 4 $ $R$-matrix. Since the later possibility 
leads us to trivial manifolds we will
disregard it, as did Kasteleyn, from our forthcoming analysis.

The probable reason that
such generalization has not yet been carried out 
seen technical since in the asymmetric model we have
to deal with the double number of equations as compared to the
symmetric eight-vertex model. At first sight this
appears to be a cumbersome task, but here we show 
that it is possible to simplify this problem,
without recoursing to computer manipulations, 
to a number of simple equations that 
will clarify the common origin of
the above two integrable manifolds.  Besides that, this approach
allows to show that
manifold (3) is one between three possible different integrable
branches satisfying the free-fermion condition.
The ``star-triangle'' relations are sufficient conditions \cite{BA} for commuting
transfer matrices and for the asymmetric
eight-vertex model they are given by
\begin{equation}
a_{\pm} a_{\pm}^{'}d^{"} +dc^{'}a_{\mp}^{"}=
cd^{'}a^{"}_{\pm}+b_{\mp}b_{\mp}^{'}d^{"}
\end{equation}
\begin{equation}
d b_{\pm}^{'}c^{"} +a_{\pm}d^{'}b_{\mp}^{"}=
b_{\pm}d^{'}a^{"}_{\pm}+cb_{\mp}^{'}d^{"}
\end{equation}
\begin{equation}
d b_{\pm}^{'}b^{"}_{\pm} +a_{\pm}d^{'}c^{"}=
da^{'}_{\pm}a^{"}_{\pm}+a_{\mp}c^{'}d^{"}
\end{equation}
\begin{equation}
ca_{\pm}^{'}c^{"} +b_{\pm}c^{'}b^{"}_{\mp}=
a_{\pm}c^{'}a^{"}_{\pm}+da_{\mp}^{'}d^{"}
\end{equation}
\begin{equation}
ca_{\pm}^{'}b^{"}_{\pm} +b_{\pm}c^{'}c^{"}=
cb^{'}_{\pm}a^{"}_{\pm}+b_{\mp}d^{'}d^{"}
\end{equation}
\begin{equation}
b_{\mp}a_{\pm}^{'}c^{"} +cc^{'}b_{\mp}^{"}=
dd^{'}b^{"}_{\pm}+a_{\pm}b^{'}_{\mp}c^{"}
\end{equation}

Note that each of these equations possesses two possibilities and
we shall
denote them by
Eqs.($6_{\pm}$,$\cdots$,$11_{\pm}$). Altogether we have twelve
linear homogeneous equations and only six weights,
say
$ a_{\pm}^{"},b_{\pm}^{"},c^{"}$ and $d^{"} $, are  at our disposal
to be eliminated in terms of remaining  set of weights
$\{ a_{\pm},b_{\pm},c,d\}$ and
$\{ a_{\pm}^{'},b_{\pm}^{'},c^{'},d^{'}\}$.  Therefore we have to choose
the appropriate equations to start with and our
solution goes as follows. 
We first eliminate the weights $a_{\pm}^{"}$ with the help of the pair
of equations ($9_{\mp}$,$10_{\pm}$) and by substituting the result in
Eqs.($6_{\pm}$) we find the following relations
\begin{eqnarray}
b_{\pm}^{"}(a_{\mp}ca_{\pm}^{'}d^{'}-db_{\mp}b_{\pm}^{'}c^{'}) & =&
c^{"}(cda_{\mp}^{'}b_{\pm}^{'}-a_{\mp}b_{\pm}c^{'}d^{'}) \nonumber\\
& &
+d^{"} \left[ a_{\mp}b_{\mp}({d^{'}}^2-b_{\pm}^{'}b_{\mp}^{'}) +
a_{\pm}^{'}b_{\pm}^{'}(a_{\pm}a_{\mp}-d^2)  \right]
\end{eqnarray}

Next we apply similar procedure in the case of Eqs.($7_{\mp}$,$8_{\pm}$)
and the corresponding relations between the weights $b_{\pm}^{"}$, $c^{"}$
and $d^{"}$ are
\begin{eqnarray}
b_{\pm}^{"}(db_{\mp}b_{\pm}^{'}c^{'}-a_{\mp}ca_{\pm}^{'}d^{'}) & =&
c^{"}(cda_{\pm}^{'}b_{\mp}^{'}-a_{\pm}b_{\mp}c^{'}d^{'}) \nonumber\\
& &
+d^{"} \left[ a_{\mp}b_{\mp}({c^{'}}^2-a_{\pm}^{'}a_{\mp}^{'}) +
a_{\pm}^{'}b_{\pm}^{'}(b_{\pm}b_{\mp}-c^2)  \right]
\end{eqnarray}

From Eqs.($12_{\pm},13_{\pm}$) it is not difficult to eliminate the
weights $b_{\pm}^{"}$, leading us to constraints between
$c^{"}$ and $d^{"}$
\begin{equation}
c^{"}\left[cd(a_{\pm}^{'}b_{\mp}^{'}+a_{\mp}^{'}b_{\pm}^{'})
-c^{'}d^{'}(a_{\pm}b_{\mp}+a_{\mp}b_{\pm}) \right]=
d^{"}\left[a_{\mp}b_{\mp}F(a_{\pm}^{'},b_{\pm}^{'},c^{'},d^{'})-a_{\pm}^{'}
b_{\pm}^{'}F(a_{\pm},b_{\pm},c,d) \right]
\end{equation}

At this point it is tempting to use such equations and the previous
results for $a_{\pm}^{"}$ and $b_{\pm}^{"}$ to eliminate five weights
ratios and to substitute them in the remaining equations, namely
Eqs.($11_{\pm}$) and either Eqs.($12_{\pm}$) or Eqs.($13_{\pm}$).
This is, however, not so illuminating because it leads us to
carry out simplifications in complicated expressions. We find that
it is more profitable to repeat the procedure described above but
now we first eliminate the weights $b^{"}_{\pm}$ and in the end
we use Eqs.($11_{\pm}$) instead of Eqs.($6_{\pm}$). This leads us
to a different constraint between $c^{"}$ and $d^{"}$ given by,
\begin{equation}
d^{"}\left[cd(a_{\pm}^{'}b_{\mp}^{'}+a_{\mp}^{'}b_{\pm}^{'})
-c^{'}d^{'}(a_{\pm}b_{\mp}+a_{\mp}b_{\pm}) \right]=
c^{"}\left[a_{\mp}b_{\mp}F(a_{\pm}^{'},b_{\pm}^{'},c^{'},d^{'})-a_{\mp}^{'}
b_{\mp}^{'}F(a_{\pm},b_{\pm},c,d) \right]
\end{equation}

Now we reached a point that enables us to make conclusions on the
way the set of weights $\{a_{\pm},b_{\pm},c,d\}$
and $\{a_{\pm}^{'},b_{\pm}^{'},c^{'},d^{'}\}$ should be related
to each other. In fact, from Eqs.($14_{\pm}$,$15_{\pm}$) we
find that the necessary conditions for the weights $c^{"}$ and
$d^{"}$ not to be all zero are
\begin{equation}
\frac{cd}{a_{+}b_{-}+a_{-}b_{+}}=
\frac{c^{'}d^{'}}{a_{+}^{'}b_{-}^{'}+a_{-}^{'}b_{+}^{'}}
\end{equation}
and either
\begin{equation}
F(a_{\pm},b_{\pm},c,d)=
F(a_{\pm}^{'},b_{\pm}^{'},c^{'},d^{'})=0
\end{equation}
or
\begin{equation}
\frac{a_{+}b_{+}}{a_{-}b_{-}}=
\frac{a_{+}^{'}b_{+}^{'}}{a_{-}^{'}b_{-}^{'}}=1,~~
\frac{F(a_{\pm},b_{\pm},c,d)}{a_{-}b_{-}}=
\frac{F(a_{\pm}^{'},b_{\pm}^{'},c^{'},d^{'})}{a_{-}^{'}b_{-}^{'}}
\end{equation}

We are already in the position to conclude that the asymmetric
eight-vertex model has indeed only two possible integrable manifolds,
one is singled out by the free-fermion condition (17) while
the other (18) turns out to be a mixed type of conditions that relate
the set of weights both alone and between each other.  
One important point of our analysis is
that it makes clear that both manifolds need to share a common invariant
given by Eq.(16)

To close our analysis it remains to check the consistency
between Eqs.($6_{\pm}$) and Eqs.($11_{\pm}$) which can in principle
be a source of further constraints. From such equations one can easily
calculate the ratios $\frac{a_{+}^{"}}{a_{-}^{"}}$
and $\frac{b_{+}^{"}}{b_{-}^{"}}$, namely
\begin{equation}
\frac{a_{+}^{"}}{a_{-}^{"}}= \frac{cd^{'}(a_{+}a_{+}^{'}-b_{-}b_{-}^{'})
-dc^{'}(b_{+}b_{+}^{'}-a_{-}a_{-}^{'})}
{dc^{'}(a_{+}a_{+}^{'}-b_{-}b_{-}^{'})
-cd^{'}(b_{+}b_{+}^{'}-a_{-}a_{-}^{'})}
\end{equation}
\begin{equation}
\frac{b_{+}^{"}}{b_{-}^{"}}= \frac{dd^{'}(b_{-}a_{+}^{'}-a_{+}b_{-}^{'})
-cc^{'}(a_{-}b_{+}^{'}-b_{+}a_{-}^{'})}
{cc^{'}(b_{-}a_{+}^{'}-a_{+}b_{-}^{'})
-dd^{'}(a_{-}b_{+}^{'}-b_{+}a_{-}^{'})}
\end{equation}
which in principle can be compared with our previous results for the
same ratios.

Before proceeding with that, however, there exists one property that
we have not yet explored. Instead of starting our analysis by 
eliminating the weights $a^{"}_{\pm}$,$b_{\pm}^{"}$,
$c^{"}$ and $d^{"}$ we could choose to begin with the other two sets
of weights as well. Because the ``star-triangle'' equations
are not symmetric by exchanging a given two sets of 
weights we expect
that each  possibility will leads us to different kind of
constraints. This means that we can use the asymmetry of the
weights in our favour which may help us in further
simplifications.  For example, the relations ($6_{\pm}$-$11_{\pm}$)
are invariant under the exchange of weights
$\{a_{\pm}^{"},b_{\pm}^{"},c^{"},d^{"}\} $  and
$\{a_{\pm},b_{\pm},c,d\} $ only after the transformation 
$b_{\pm} \rightarrow b_{\mp}$ is performed for all set of weights.
This means that if we had started our procedure by
eliminating the weights $a_{\pm},b_{\pm},c$ and $d$ the 
same analysis we have carried out so far will lead us to the following
constraints
\begin{equation}
\frac{c^{'}d^{'}}{a_{+}^{'}b_{+}^{'}+a_{-}^{'}b_{-}^{'}}=
\frac{c^{"}d^{"}}{a_{+}^{"}b_{+}^{"}+a_{-}^{"}b_{-}^{"}}
\end{equation}
besides that either
\begin{equation}
F(a_{\pm}^{'},b_{\pm}^{'},c^{'},d^{'})=
F(a_{\pm}^{"},b_{\pm}^{"},c^{"},d^{"})=0
\end{equation}
or
\begin{equation}
\frac{a_{-}^{'}b_{+}^{'}}{a_{+}^{'}b_{-}^{'}}=
\frac{a_{-}^{"}b_{+}^{"}}{a_{+}^{"}b_{-}^{"}}=1,~~
\frac{F(a_{\pm}^{'},b_{\pm}^{'},c^{'},d^{'})}{a_{-}^{'}b_{+}^{'}}=
\frac{F(a_{\pm}^{"},b_{\pm}^{"},c^{"},d^{"})}{a_{-}^{"}b_{+}^{"}}
\end{equation}

By the same token if we had started by eliminating 
$a_{\pm}^{'},b_{\pm}^{'},c^{'}$ and $d^{'}$ we will find
\begin{equation}
\frac{cd}{a_{+}b_{+}+a_{-}b_{-}}=
\frac{c^{"}d^{"}}{a_{+}^{"}b_{-}^{"}+a_{-}^{"}b_{+}^{"}}
\end{equation}
and that either
\begin{equation}
F(a_{\pm},b_{\pm},c,d)=
F(a_{\pm}^{"},b_{\pm}^{"},c^{"},d^{"})
\end{equation}
or
\begin{equation}
\frac{a_{-}b_{+}}{a_{+}b_{-}}=
\frac{a_{+}^{"}b_{+}^{"}}{a_{-}^{"}b_{-}^{"}}=1,~~
\frac{F(a_{\pm},b_{\pm},c,d)}{a_{-}b_{+}}=
\frac{F(a_{\pm}^{"},b_{\pm}^{"},c^{"},d^{"})}{a_{-}^{"}b_{-}^{"}}
\end{equation}

Let us now analyze the consequences of this observation for each possible
integrable manifold and here we begin with
the second manifold. It is not difficult to see that the
consistency of the equations (18),(23) and (26),  to what concern
relations within the same set of weights, impose severe restrictions on
the second type of the manifold, namely 
\begin{equation}
a_{+}=a_{-}~~and~~b_{+}=b_{-}~~~~or~~~~
a_{+}=-a_{-}~~and~~b_{+}=-b_{-}
\end{equation}
and similar conditions for the other sets $\{a_{\pm}^{'},b_{\pm}^{'}\}$
and $\{a_{\pm}^{"},b_{\pm}^{"}\}$.

It turns out, however, that the only possibility compatible with
the ``universal'' constraints (16), (21) and (24) is the totally
symmetric case $a_{+}=a_{-}$ and $b_{+}=b_{-}$ leading 
us therefore to the Baxter's model (5). Note that in this situation the
compatibility between Eq($6_{\pm}$) and Eq.($11_{\pm}$)  is trivial
because both equations (19) and (20) are automatically
satisfied.

We now turn our attention to the free-fermion
manifold. In this case   we have much less
restrictive constraints since we are only left with relations between different
weights, namely Eqs.(16,21,24). Altogether these equations provide us
the following relation  
\begin{equation}
\frac{a_{+}^{"}b_{+}^{"}+a_{-}^{"}b_{-}^{"}}
{a_{+}^{"}b_{-}^{"}+a_{-}^{"}b_{+}^{"}}=
\frac{a_{+}b_{-}+a_{-}b_{+}}
{a_{+}b_{+}+a_{-}b_{-}}
\frac{a_{+}^{'}b_{+}^{'}+a_{-}^{'}b_{-}^{'}}
{a_{+}^{'}b_{-}^{'}+a_{-}^{'}b_{+}^{'}}
\end{equation}
whose compatibility with 
Eqs.($6_{\pm}$,$11_{\pm}$) can be implemented by evaluating
the left-hand side of Eq.(28) with the help of Eqs.(19,20). After
few manipulations, in which the free-fermion condition
is explicitly used, we end up with
a ``separable'' equation $P =0$ for the weights
$\{a_{\pm},b_{\pm},c,d\} $  and
$\{a_{\pm}^{'},b_{\pm}^{'},c^{'},d^{'}\} $ 
and the polynomial $P$ is given by
\begin{eqnarray}
P &=& \left [ (c^2+d^2)(a_{+}^{'}b_{+}^{'}+a_{-}^{'}b_{-}^{'})
-(a_{+}b_{+}+a_{-}b_{-})({c^{'}}^{2}+{d^{'}}^{2}) \right ]
\nonumber\\
&&
\times \left [ a_{-}b_{+}a_{-}^{'}b_{+}^{'}-a_{+}b_{-}a_{+}^{'}b_{-}^{'} 
\right ] \nonumber\\
&&
\times
\left [ (a_{+}^{2}+b_{-}^2-a_{-}^2-b_{+}^2)(a_{+}^{'}b_{-}^{'}
+a_{-}^{'}b_{+}^{'}) -
(a_{+}b_{-}
+a_{-}b_{+})
({a_{+}^{'}}^{2}+{b_{-}^{'}}^2-{a_{-}^{'}}^2-{b_{+}^{'}}^2)
\right ]  \nonumber\\
\end{eqnarray}

From this equation  we conclude that we have three possible free-fermion
integrable manifolds given by either
\begin{equation}
\frac{a_{+}^{2}+b_{-}^{2}-a_{-}^2 -b_{+}^{2}}{a_{+}b_{-}+a_{-}b_{+}}=
\frac{{a_{+}^{'}}^{2}+{b_{-}^{'}}^{2}-{a_{-}^{'}}^2 -{b_{+}^{'}}^{2}}{a_{+}^{'}b_{-}^{'}+a_{-}^{'}b_{+}^{'}}
\end{equation}
or
\begin{equation}
\frac{a_{+}b_{-}}{a_{-}b_{+}}=
\frac{a_{+}^{'}b_{-}^{'}}{a_{-}^{'}b_{+}^{'}}=\pm 1
\end{equation}
or still
\begin{equation}
\frac{c^2+d^2}{a_{+}b_{+}+a_{-}b_{-}}=
\frac{{c^{'}}^2+{d^{'}}^2}{a_{+}^{'}b_{+}^{'}+a_{-}^{'}b_{-}^{'}}
\end{equation}
besides of course the free-fermion conditions for both 
$\{a_{\pm},b_{\pm},c,d\} $  and
$\{a_{\pm}^{'},b_{\pm}^{'},c^{'},d^{'} \} $ together with
the ``universal'' relation (16).
Note that the free-fermion case (31) can not be related to the
manifold (4) beginning by the fact that in the former model the weight $c$ can be different of
the weight $d$.

In Figure 1 we have summarized all the results obtained so far. Let us
now compare our results with previous work in the literature. 
Contrary to what happened to the symmetric manifold (5)
we recall
that Eqs.(30,31) do not imply that the ratios
$\frac{a_{+}^{2}+b_{-}^{2}-a_{-}^2 -b_{+}^{2}}{a_{+}b_{-}+a_{-}b_{+}}$
and 
$\frac{c^2+d^2}{a_{+}b_{+}+a_{-}b_{-}}$ are 
necessarily constants but only that
they are invariants for two distinct set of weights \footnote{
The same statement is of course valid for the ``universal''
ratio $\frac{cd}{a_{+}b_{-}+a_{-}b_{+}}$.}. This is the
reason why general solutions of the Yang-Baxter equation satisfying
the free-fermion condition are expected to be non-additive \cite{CH}.
In fact, in the appendix $A$ we show that the additional
assumption of additivity provides us extra restriction to the weights.
In this sense, the manifold (30) turns out to be a generalization
of the original result (3) by Krinsky \cite{KI}. Next 
the manifold (31) has been only partially obtained
in the literature, more precisely in the special case
$a_{+}=a_{-}$ and $b_{+}=-b_{-}$
\cite{WA,CH}\footnote{ Of course the
other possibility $a_{+}=a_{-}$ and
$b_{+}=b_{-}$ is contained in the Baxter's solution.}. 
Finally, to the best of our knowledge the last branch (32)
is new in the literature.
The probable reason
that such general manifolds have been missed in previous work, 
see for example refs.\cite{WA,CH},
is related to analysis of the Yang-Baxter equation 
in terms of spectral parameters. 
There it was required that at certain value
of the spectral parameter (initial condition) the weights should be regular, i.e that
the corresponding $L_j$ operator be proportional to the four dimensional permutator.
Note that the $L_j$ operator of manifold (32) can not be made regular and therefore does not
have a local associated Hamiltonian. This is also the reason Barouch \cite {BO} and
Kasteleyn \cite{KA} missed such manifold since they used the
assumption of local forms of Hamiltonians.
We recall that though the property of regularity 
guarantees that the logarithmic derivative of the transfer matrix
is $local$ this is by no means a necessary 
condition for integrability.

In summary, we have analyzed according to Baxter the integrable 
branches of the asymmetric eight-vertex model. Besides 
recovering the Baxter's model we shown that the free-fermion
condition produces three different set of integrable manifolds. 
A natural question to be asked is whether or not the new
manifolds (31) and (32) can be solved by the method devised
by Felderhof originally proposed to diagonalize the transfer
matrix of the Krinsky's manifold (3). This is of interest 
since these systems maybe the corner stone of highly non trivial
models as recently have been discussed in refs.\cite{CH1,CH2}. In fact,
we have evidences that the manifold (31)
is related to a staggered $XY$ model. 
Because both the Baxter symmetric model
and the free-fermion manifolds (30-32) share 
a common algebraic structure,
the Yang-Baxter algebra, it is plausible to think that
Baxter's generalized Bethe ansatz can be adapted to include
the solution of the free-fermion models too.  This problem
has eluded us so far though some progress has been made in
the case of the simplest free-fermion branch (31).

\centerline{\bf Appendix A }
\setcounter{equation}{0}
\renewcommand{\theequation}{A.\arabic{equation}}

The purpose here is to demonstrate that the
hypothesis of additivity of the weights leads us to
much more restrictive conditions for the free-fermion
manifold as compared with the results (30-32) of the main text.
In order to see that lets us consider as usual 
that the weights 
${a_{\pm},b_{\pm},c,d}$ are parameterized by the variables
$x_1$ and $x_2$  and similarly 
that ${a_{\pm}^{'},b_{\pm}^{'},c^{'},d^{'}}$ 
and ${a_{\pm}^{"},b_{\pm}^{"},c^{"},d^{"}}$  are parameterized by $x_1,x_3$
and $x_2,x_3$, respectively.  The consistency between the ``universal''
relations (16), (21) and (24) implies  a remarkable 
separability condition for the ratio
\begin{equation}
\frac{a_{+}(x_1,x_2)b_{-}(x_1,x_2) +a_{-}(x_1,x_2)b_{+}(x_1,x_2)}
{a_{+}(x_1,x_2)b_{+}(x_1,x_2) +a_{-}(x_1,x_2)b_{-}(x_1,x_2)}=\frac{G(x_1)}{G(x_2)}
\end{equation}
where $G(x)$ is an arbitrary function.

The additional assumption that the weights are additive means that this
function is necessarily a constant which ultimately leads us to the
relation
\begin{equation}
(a_{+}-a_{-})(b_{+}-b_{-})=0
\end{equation}

As a consequence of that the possible manifolds  satisfying the 
the free-fermion condition are either $a_{+}=a_{-}$ or
$b_{+}=b_{-}$. Now by imposing the consistency between Eqs($6_{\pm}$)
and Eqs.($11_{\pm}$) it turns out that these two possibilities becomes
either
\begin{equation}
a_{+}=a_{-}~~~and~~~b_{+}=-b_{-}
\end{equation}
or
\begin{equation}
b_{+}=b_{-},~~~and~~~
\frac{a_{+}-a_{-}}{b_{+}}=\Delta
\end{equation}
where $\Delta$ is a constant.

Clearly, these are special cases of the manifolds (31) and (30),
respectively.

\section*{Acknowledgements} 
This work has been partially supported
by the Brazilian research Agencies CNPq and Fapesp.

\newpage
\begin{figure}[t]
\centerline{\epsfxsize=\textwidth\epsfbox{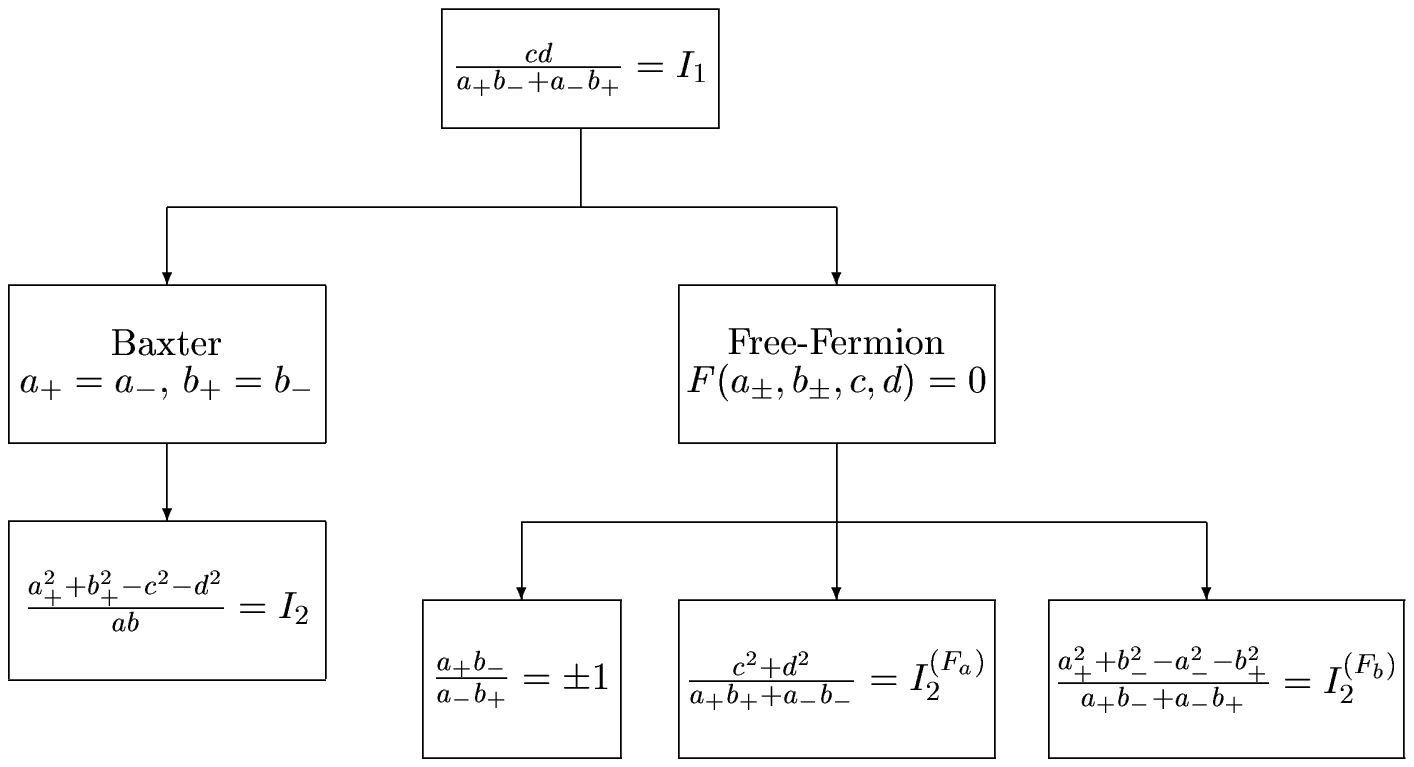}}
\vskip -12cm
\caption{ 
\protect
{\footnotesize 
Summary of the integrable manifolds of the asymmetric eight-vertex model.
The symbols $I_1$, $I_2^{(F_a)}$ and $I_2^{(F_b)}$ denote invariants for two
distinct set of weights.
}}
\end{figure}

\end{document}